\documentclass[aps,twocolumn,pra]{revtex4}

\usepackage{amsmath,amssymb,bm}
\usepackage{graphicx}
\usepackage{epstopdf}
\usepackage{latexsym}
\usepackage{subfigure}
\usepackage[usenames,dvipsnames]{color}
\usepackage{hyperref}
\usepackage{natbib}
\usepackage{moreverb}
\usepackage{xparse}

\begin{document}
\newcommand{\cs}[1]{{\color{blue}$\clubsuit$#1}}
\newcommand{\DD}[2]{\left(2-\delta_{\mathbf{n}_{#1}\mathbf{n}_{#2}}\right)}
\newcommand{\DE}[2]{2\left(1-\delta_{\mathbf{n}_{#1}\mathbf{n}_{#2}}\right)}
\NewDocumentCommand{\dE}{mmg}{\Delta E_{\mathbf{n}_{#1}}\left(q_{#2},\IfNoValueTF{#3}{\phi}{#3}\right)}
\newcommand{\V}[3]{V^{#1}_{\{\mathbf{n}_{#2}\mathbf{n}_{#3}\}}}
\newcommand{\U}[3]{U^{#1}_{\{\mathbf{n}_{#2}\mathbf{n}_{#3}\}}}
\newcommand{\sinc}{\mathrm{sinc}}
\newcommand{\red}[1]{{ \bf \color{red}#1}}

\title{ Synthetic spin-orbit coupling in an  optical lattice clock}
\author{Michael L.~Wall$^1$, Andrew P.~Koller$^1$, Shuming Li$^1$, Xibo Zhang$^1$, Nigel R. Cooper$^2$, Jun Ye$^1$, and Ana Maria Rey$^1$}
\affiliation{$^1$JILA, NIST, Department of Physics, University of Colorado, 440 UCB, Boulder, CO 80309, USA}
\affiliation{$^2$T.C.M. Group, Cavendish Laboratory, J.J. Thomson Avenue, Cambridge CB3 0HE, United Kingdom}
\date{\today}

\begin{abstract}
We propose the use of optical lattice clocks operated with  fermionic alkaline-earth-atoms to study spin-orbit coupling (SOC)  in interacting many-body systems. The SOC emerges naturally during the clock  interrogation when  atoms  are allowed to tunnel and accumulate a phase set by the ratio of the ``magic'' lattice wavelength to the clock transition wavelength.  We  demonstrate how  standard  protocols such as Rabi and Ramsey spectroscopy, that  take advantage of the sub-Hertz resolution of   state-of-the-art  clock lasers, can perform momentum-resolved band  tomography and  determine  SOC-induced $s$-wave collisions in nuclear spin polarized  fermions.  By adding a second counter-propagating clock beam a sliding superlattice can be implemented  and used for controlled atom transport and as a probe of $p$ and  $s$-wave interactions. The proposed spectroscopic probes provide clean and well-resolved signatures at current clock operating temperatures.
\end{abstract}
\date{\today}
\maketitle

The recent implementation  of synthetic gauge fields  and spin-orbit coupling  (SOC) in neutral atomic gases ~\cite{Galitski2013,NigelCooperReview,Dalibard,Spielmanreview,Zhang2015,Stuhl_spielman,Mancini_Fallani} is a groundbreaking step towards the ultimate goal of using these fully-controllable systems to synthesize and probe novel topological states of matter.
So far  optical Raman transitions have been used to couple different internal (e.g.~hyperfine) states while transferring net momentum to the atoms. However, in alkali atoms Raman-induced spin-flips inevitably suffer from  heating mechanisms associated with spontaneous emission. While this issue has not yet been an impediment for the investigation of  non-interacting processes or mean field effects~\cite{Spielmanreview,Zhang2015}, it could limit the ability to observe interacting many-body phenomena that manifest at longer timescales. Finding alternative, more resilient methods for generating synthetic SOC, and  probing its  interplay with  interactions, is thus highly desirable.

To reduce heating, the  use of atoms with richer internal structure such as alkaline-earth-atoms (AEAs) \cite{Dalibard,Mancini_Fallani} or   lanthanide atoms such as Dy and Er \cite{Hui2013} have been suggested. Here, we propose an implementation of SOC  using  cold AEAs in  an optical lattice clock (OLC)~\cite{OLCs} and discuss how  SOC physics can be probed in interacting many-body systems. Our proposal relies on the fact that SOC naturally arises in OLCs because the clock laser imprints  a phase that varies significantly from one lattice site to the next as it drives  an   ultra-narrow optical transition.    Our proposal offers several advantages over prior schemes. It  uses a direct transition to a long-lived  electronic clock state with natural lifetime $\gtrsim10^2$ s~\cite{Boyd2006}, and  thus heating from spontaneous emission is negligible. Second, our proposal takes advantage of the sub-Hz resolution of clock lasers~\cite{Bloom,Nicholson,Hinkley,Katori}.  Finally, we can probe the interplay of interactions and SOC by operating in the  regime  where the interaction energy per particle, $U$,  is weak compared to the characteristic trapping energies ~\cite{Martin_2013,lemke2011,SUN2,Rey_Annals,losses1,losses2} but comparable to SOC scales determined by $J$, the tunneling, and $\Omega$, the clock Rabi frequency.

\begin{figure}[t]
\centering
\includegraphics[width=\columnwidth]{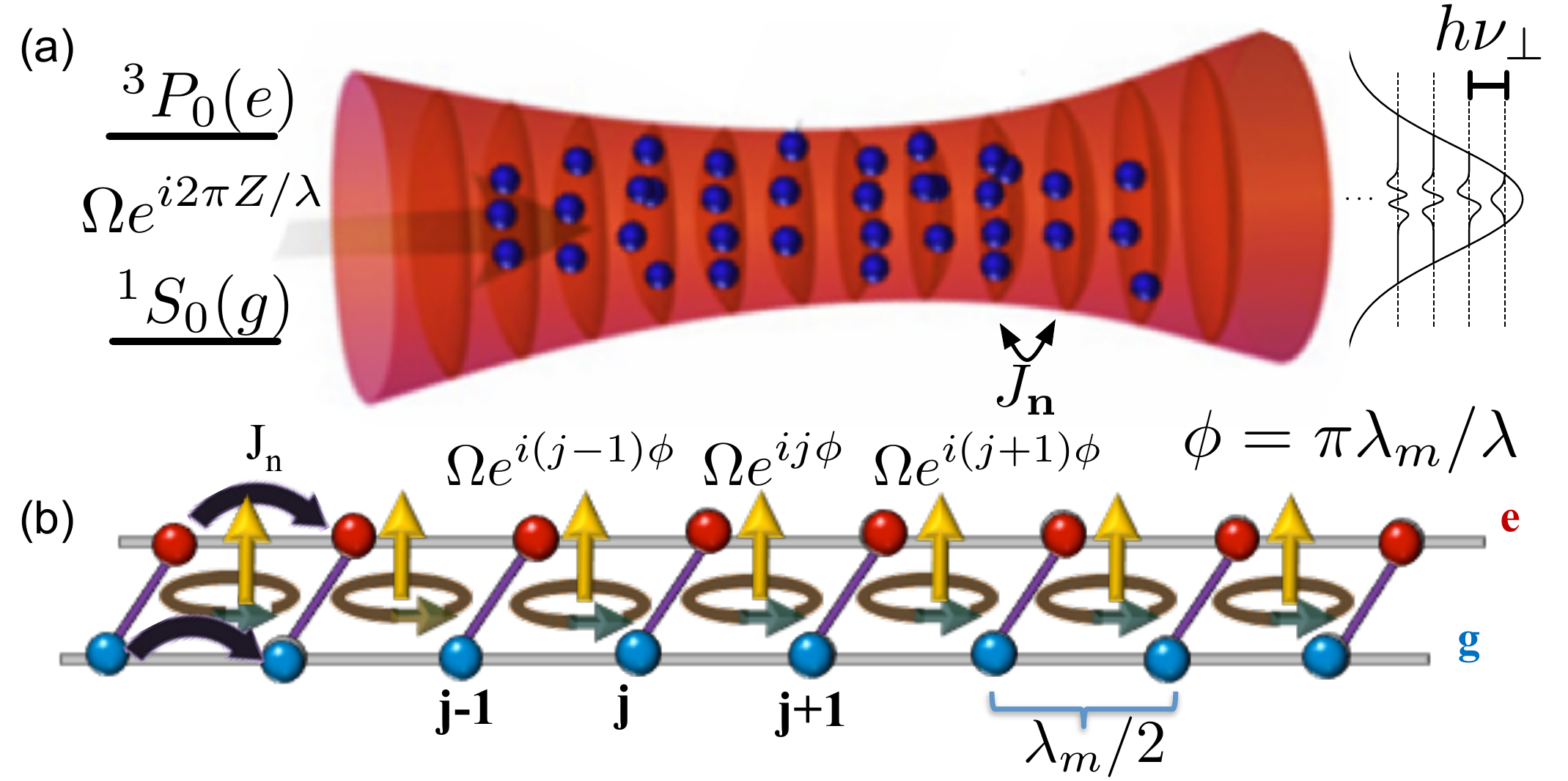}
\caption{(Color online) (a) A clock laser along the $Z$ direction of wavelength $\lambda$ and Rabi frequency $\Omega$ interrogates the $^1S_0(g)$-$^3P_0(e)$ transition in fermionic alkaline-earth atoms  trapped in an optical lattice with magic wavelength $\lambda_{m}$.  The transverse confinement is provided by Gaussian curvature of the lattice beams with harmonic frequency $h\nu_{\perp}$. Many transverse modes $\mathbf{n}$ are populated at current operating temperatures. (b) The phase difference $\phi$ between adjacent sites $j$ and $(j+1)$ induces  SOC when atoms can tunnel with mode-dependent tunnel-coupling $J_{\mathbf{n}}$, realizing a synthetic two-leg ladder with flux $\phi$ per plaquette.  }
\label{fig:OLC}
\end{figure}

We propose three protocols for probing  SOC in an OLC and demonstrate that all provide clear signatures under current operating conditions.  In the first protocol, we demonstrate that momentum-resolved tomography of chiral band structures can be performed using Rabi spectroscopy in the parameter regime $J\sim\Omega\gg U$.  In the second protocol we show that the modification of collisional properties by SOC~\cite{Williams,Zhang} is manifest in standard Ramsey spectroscopy, focusing on the parameter regime $J\gtrsim U$.  In the final protocol, controlled and spatially resolved atomic transport~\cite{Thouless,TroyerPump,KwekPump,MuellerPump,CooperRey,Blochpump,Takahasipump,Spielmanpump} is induced by an additional counter-propagating clock beam that exhibits  a controllable phase difference with respect to the original probe beam, and interaction effects beyond mean field modify the dynamics in the parameter regime $\Omega\gg J\sim U$.  Although  inelastic collisions in the excited state impose limitations in the probing time \cite{losses1,losses2,Scazza,Cappelinni}, we also show that they can be used as a resource for state preparation and readout.



{\it SOC implementation}  Current OLCs interrogate  the $^1S_0(g)$-$^3P_0(e)$ transition of ensembles  of thousands of  nuclear spin polarized fermionic AEAs  trapped in a deep 1D  optical lattice  that  splits the gas  in  arrays of 2D pancakes~\cite{OLCs}(see Fig.\ref{fig:OLC}). The lattice potential uses the   magic-wavelength, $\lambda_{m}$ to generate identical trapping conditions  for the two states.   At  current operating temperatures,  $T \sim \mu$K~\cite{Martin_2013} the population of higher axial  bands  is negligible ($\lesssim 5\%$).  On the other hand, along the transverse directions, where  the confinement is  provided wholly by the Gaussian curvature of the optical lattice  beams, modes are thermally populated with an average number of mode quanta $\langle n\rangle\sim 50$.  To generate SOC coherent tunneling between lattice sites is required. Our proposal is to superimpose a running-wave beam on the lattice potential:
\begin{align}
\textstyle V_{\mathrm{ext}}\left(\mathbf{R}\right)&=\textstyle-\exp\left(-\frac{2R^2}{w_0^2}\right)\left[V_{\mathrm{const}}+V\cos^2\left(\frac{2\pi Z}{\lambda_m}\right)\right]\, ;
\label{eq:LattPot}
\end{align}
this increases the transverse confinement without significantly affecting the axial motion~\cite{Wall_Hazzard_15}.  Here, $w_0$ is the beam waist, $R$ the transverse radial coordinate, $Z$ the axial coordinate, $V$ the axial lattice corrugation, and $V_{\mathrm{const}}$ the running-wave induced  potential.  By increasing $V_{\mathrm{const}}$ as $V$ is lowered, the transverse confinement frequency $\nu_\perp \sim \sqrt{\frac{1 }{m\pi^2 w_0^2}(V_{\mathrm{const}}+V)}$
is kept constant while the tunneling rate along the axial direction increases.  Here $m$ is the atom mass.

Since $V_{\mathrm{ext}}\left(\mathbf{R}\right)$ is discretely translationally invariant along the axial direction, atoms   trapped in the lowest axial lattice band are   governed  by the Hamiltonian $\hat{H}^0=\sum_\mathbf{n} \hat{H}^0_\mathbf{n}$~\cite{Supp}
\begin{eqnarray}\label{eq:Hsprs}
\hat{H}^0_\mathbf{n}=\sum_{q,\alpha} E_{\alpha,\mathbf{n},q}  \hat{n}_{\alpha,\mathbf{n},q} - \sum_{q}\left[\frac{\Omega_{\mathbf{n}}}{2}\hat{a}^{\dagger}_{+,\mathbf{n},q+\phi}\hat{a}_{-,\mathbf{n},q}+\mathrm{H.c.}\right]\, .
 \end{eqnarray} Here and throughout $q=\tilde{q}a$ is the dimensionless product of axial quasimomentum $\tilde{q}$ and lattice spacing $a=\lambda_m/2$, $\hat{a}_{\alpha ,\mathbf{n},q}$ annihilates  a fermion in  the two-dimensional transverse mode $\mathbf{n}$, quasimomentum $q$,  state  $\alpha=\pm$(for $e$ and $g$),  and $\hat{n}_{\alpha,\mathbf{n},q}=\hat{a}_{\alpha,\mathbf{n},q}^\dagger \hat{a}_{\alpha,\mathbf{n},q} $.   The energy  $E_{ \alpha,\mathbf{n},q}\left(q\right)=\alpha\frac{\delta}{2}  +\bar{E}_{\mathbf{n}}-2J_\mathbf{n} \cos(q)$ has contributions from   the mode dependent tunneling $J_\mathbf{n} $, the average energy of the transverse mode $\mathbf{n}$, $\bar{E}_{ \mathbf{n}}$, and   the laser detuning  $\delta$.  $\Omega_{\mathbf{n}}$ is the Rabi frequency  for  mode $\mathbf{n}$. The clock laser with wavelength $\lambda$ imprints a phase that varies  between adjacent lattice sites by  $\phi=\pi\lambda_m/\lambda$.

\begin{figure}
\centering
\includegraphics[width=\columnwidth]{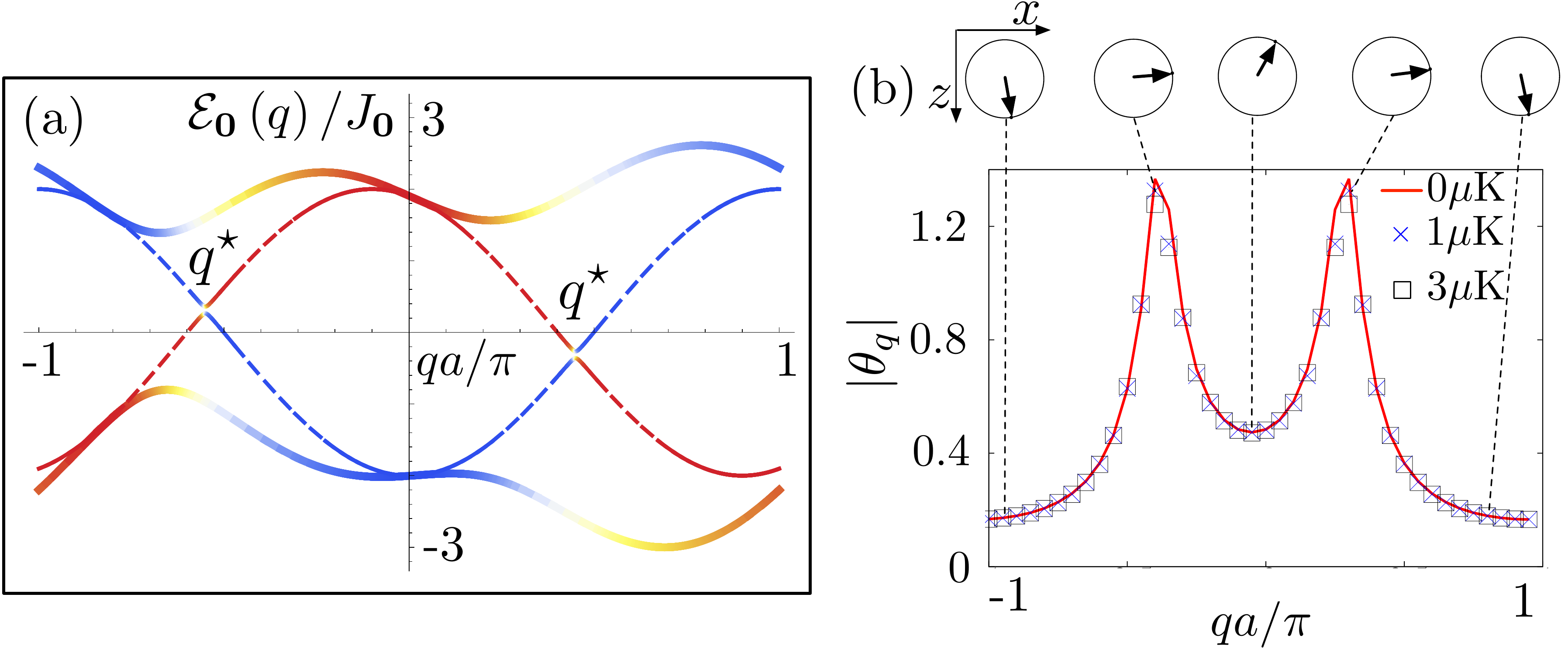}
\caption{(Color online)  (a)  SOC band structure for $\delta=-2J_{\mathbf{0}}$, $\Omega_{\mathbf{0}}=J_{\mathbf{0}}$ (solid lines) and  $\Omega_{\mathbf{0}}^p=0.05J_{\mathbf{0}}$  (dashed lines).  The axial depth is $V/E_R=12$,  $J_{\mathbf{0}}=42$ Hz  and  $h\nu_\perp\approx 900$ Hz.  Colors correspond to state character, with $g$ ($e$) being more blue (red).  (b) Chiral Bloch vector angle, $\theta_{\mathbf{0}q}$,  in the $xz$ plane  extracted from Rabi spectroscopy using the protocol explained in the text~\cite{Supp}. The figure shows three temperatures for the parameters of (a).
}
\label{fig:2}
\end{figure}

If ones views the two internal states as a discrete synthetic  dimension~\cite{Celi,Paredes},  as shown in Fig.~\ref{fig:OLC}, $\hat{H}^0_\mathbf{n}$  describes the motion of a charged particle on a two-leg ladder  in a magnetic field with  flux $\phi$ per lattice plaquette.  $\hat{H}^0$ hence has the interpretation of many copies of those ladders -- one for each transverse mode.
By performing a gauge transformation $ \hat{a}_{+,q+\phi,\mathbf{n}} \to \hat{a}_{+,q,\mathbf{n}}$,   $\hat{H}_\mathbf{n}^0$ becomes diagonal in momentum space with  the excited state   dispersion shifted by $\phi$, $q\to q +\phi$. The latter can be then conveniently written in terms of  spin-$1/2$ operators acting on the populated modes
\begin{align}\label{eq:Bsprs}
 \hat{H}^0&=-\sum_{\mathbf{n},q} {\bf B}_{\mathbf{n}q}\cdot \hat{\vec{S}}_{\mathbf{n}q}
\end{align} where $\mathbf{B}_{\mathbf{n} q}=\left(\Omega_{\mathbf{n}},0,\Delta E_{\mathbf{n}}\left(q,\phi\right)+\delta\right)$,
$\Delta E_{\mathbf{n}}\left(q,\phi\right)\equiv 2J_{\mathbf{n}}\left[\cos\left(q\right)-\cos\left(q+\phi\right)\right]$, and
$\hat{S}^{x,y,z}_{\mathbf{n} q}$ are spin-$1/2$ angular momentum operators.
The eigenstates are described by Bloch vectors pointing in the $xz$ plane, with a  direction specified by a single angle $\theta_{\mathbf{n} q}=\arctan\left(\frac{\Omega_{\mathbf{n} }}{\Delta E_{\mathbf{n}}\left(q,\phi\right)+\delta}\right)$. The $q$ dependence of  this angle is a manifestation of chiral spin-momentum locking, which is directly connected to the topological chiral edge modes of the two-dimensional Harper-Hofstadter model~\cite{Paredes,Hatsugai}.  An example of the eigenstates and their dispersion for fixed transverse mode index is given in Fig.~\ref{fig:2}(a).


{\it Rabi spectroscopy:}  We now describe a spectroscopic protocol to probe the non-interacting chiral band structure of  Eq.~\eqref{eq:Bsprs}. In the limit of very weak Rabi frequency, $\Omega_{\mathbf{0}}\ll J_{\mathbf{0}}$, the band structure  is almost the ``bare'' one with $\Omega=0$, up to a displacement of the excited state band by $\phi$. Because of the displacement, there are special quasimomentum points $q^{\star}_{\mathbf{n}}$ where the two dispersions cross, see Fig.~\ref{fig:2}(a). These exist for a finite window of $\delta$ and are signaled in the  carrier linewidth. The width of the window is $8J_{\mathbf{n}}\left|\sin\frac{\phi}{2}\right|$ and thus  when resolved it can be used to determine $\phi$.  At finite temperature, many transverse modes are populated and hence the dependence of $\Omega_{\mathbf{n}}$ and $J_{\mathbf{n}}$ on $\mathbf{n}$ could  broaden the line and in general prevent an accurate determination of $\phi$. However, a direct simulation of the Rabi lineshape using the potential Eq.~\eqref{eq:LattPot} demonstrates that the features of the ideal, zero-temperature lineshape are captured even for a temperature of $3\mu$K~\cite{Supp}.

The ability to resolve the  $q^{\star}_{\mathbf{n}}$ resonances that appear across  the entire Brillouin zone (BZ) as the  detuning is varied can be used to perform  momentum resolved  spectroscopy and  to   precisely determine  the  chiral Bloch vector angle $\theta_{\mathbf{0}q}$ for  given values of $\Omega_{\mathbf{0}}$ and $\delta$ from Rabi oscillations. We propose the use of a three spectroscopic sequences~\cite{Supp}. One sequence  selectively excites atoms at  $q^{\star}_{\mathbf{n}}$ from $g$ to $e$ and induces Rabi oscillations. Another  filters  the dynamics of the excited atoms  from the remaining $g$ atoms  and the third is used to ``correct'' imperfections arising from  finite temperature.  As shown in Fig.~\ref{fig:2}(b), this protocol allows us to extract  $\theta_{\mathbf{0}q}$ over the BZ.  This method  is not restricted to OLCs, requiring only a stable probe, and complements other techniques for measuring band structures~\cite{Tarruell,Atala,Duca} used in  degenerate Fermi gases.

{\it Ramsey spectroscopy}  The standard Ramsey protocol starts with all atoms in the $g$ state and subjects them  to two strong, near-resonant laser pulses separated by a free evolution ``dark" time $\tau$.  The first pulse   rotates the   Bloch vector by an angle $\theta_1$ set by the pulse area, and the second converts the accumulated phase during the dark time into a $g-e$ population difference measured as Ramsey fringes. Interactions induce a density-dependent frequency shift in the fringes.

Interactions between two nuclear spin polarized atoms depend on
the motional and  electronic  degrees of freedom~\cite{Rey_Annals}. When the atoms collide they  experience $s$-wave interactions, characterized by the elastic scattering length $ a_{eg}$,  when their electronic state is antisymmetric $(|eg\rangle-|ge\rangle)/\sqrt{2}$. They can  also collide
via $p$-wave interactions, described by the corresponding  $p$-wave elastic
scattering volumes $b_{gg}^3$, $b_{ee}^3$, and $b_{eg}^3$,  in the three possible symmetric electronic configurations $|gg\rangle,|ee\rangle, (|eg\rangle+|ge\rangle)/\sqrt{2}$, respectively.  In addition to elastic interactions,  atoms can also exhibit inelastic collisions.  In $^{87}$Sr  only the $ee$ type has been observed to give rise to measurable losses~\cite{SUN2} while  in $^{173,171}$ Yb, both $ee$ and $eg$ losses have been reported~\cite{losses1,losses2,Scazza}. The effect of losses can be compensated by tracking the population decay during the dark time \cite{Martin_2013,lemke2011}.

\begin{figure}[t]
\centering
\includegraphics[width=0.66\columnwidth]{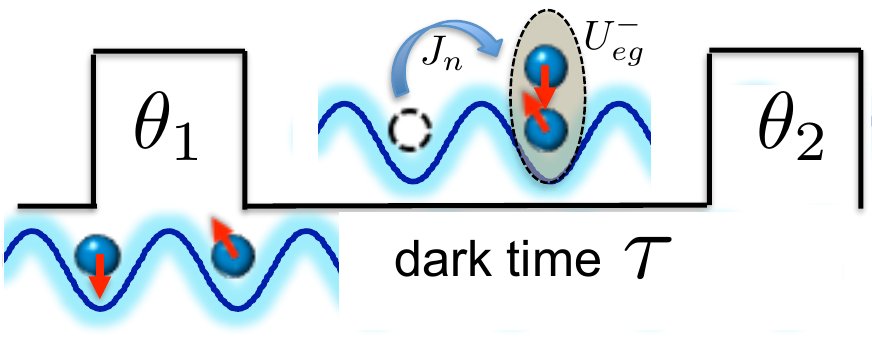}
\caption{(Color online)  A clock laser pulse with pulse area $\theta_1$ imprints a phase difference $\phi$ between atoms in neighboring sites.   Atom tunneling, $J_{\mathbf{n}}$, allows for $s$-wave interactions, $\propto U_{eg}^-$, which are signaled  as a density shift in Ramsey spectroscopy after  a second pulse of area $\theta_2$ is applied.}
\label{fig:RMSY}
\end{figure}

 When tunneling is suppressed the differential phase imparted by the laser is irrelevant, and  as long as  $\Omega_{\mathbf{n}}$ is the same for all modes -- a condition well-satisfied in current OLCs \cite{Martin_2013} --  the collective spin of  the atoms within each lattice site remains fully symmetric after the pulse and  only $p$-wave collisions occur during the dark time. Measurements under this condition ~\cite{Martin_2013,lemke2011} indeed observed a  frequency shift linearly dependent on the excitation fraction of atoms, $(1-\cos\theta_1)/2$ and fully consistent with a $p$-wave interacting model~\cite{Rey_Annals}. If instead tunneling is allowed during the dark time, atoms  become sensitive to the spatially inhomogeneous spin rotation from the site-dependent laser phase, which in turn allows for  $s$-wave collisions after a tunneling event (see Fig.~\ref{fig:RMSY}).  The  $s$-wave collisions in SOC-coupled spin polarized fermions can lead to exotic phases of matter  including  topological quantum liquids~\cite{ZhangTewari, Julia-Diaz}.


In typical OLCs  the interaction energy per particle (at the Hz level) is weaker  than the vibrational energy spacing, and large anharmonicity prevents resonant interaction-induced mode transverse changes, and so a description in terms of fixed transverse motional levels is justified~\cite{Rey_Annals}.  In the regime of weak interactions compared to tunneling, we compute the dynamics perturbatively, and find that  SOC manifests itself in the density shift at short times as~\cite{Supp}
\begin{align}
\label{eq:FOFreqShift} \Delta \nu&= \Delta \nu_0 \left[1+\frac{4\langle {J^2}\rangle_{T_R} \tau^2{\zeta}\cos\theta_1\sin^2\frac{\phi}{2}}{3\hbar^2({C}-{\chi}\cos\theta_1)}\right]\,,
\end{align} where $\Delta \nu_0=N(C-\chi \cos\theta_1 )$  is the density shift in the absence of tunneling~\cite{Martin_2013,Rey_Annals}, $N$ the mean atom number per pancake, $\langle J^2\rangle_{T_R}$ the thermally averaged squared tunneling rate, $\zeta=\left(V^{eg}-U^{eg}\right)/2$, $\chi=(V^{ee}+V^{gg}-2V^{ge})/{2}$, and $C=(V^{ee}-V^{gg})/{2}$, with $V^{\alpha\alpha'}= b_{\alpha\alpha'}^3 \langle P\rangle_{T_R}$ and $U^{eg}=a^−_{eg} \langle S\rangle_{T_R} $.
  Here  $\langle P\rangle_{T_R}\propto (T_R)^0$ and $\langle S\rangle_{T_R}\propto T_R^{-1}$ correspond to the thermal
averages of the $p$-wave and $s$-wave mode overlap coefficients respectively~\cite{Rey_Annals}, and $T_R$ is the radial temperature.  For the JILA ${}^{87}$Sr clock operated at $T_R\sim (1-5) \mu$K and $\theta_1\ll \pi$  for $\tau\sim 80$ ms, $\Delta \nu_0\sim -5$Hz. Here, $E_R$ is the recoil energy. Since SOC introduces contributions from  $s$-wave interactions, which can be one order of magnitude  larger than $p-$wave at $T_R\sim 1 \mu$K, then  from Eq. \ref{eq:FOFreqShift} we expect significant  modifications of the density shift.



\begin{figure}[t]
\centering
\includegraphics[width=\columnwidth]{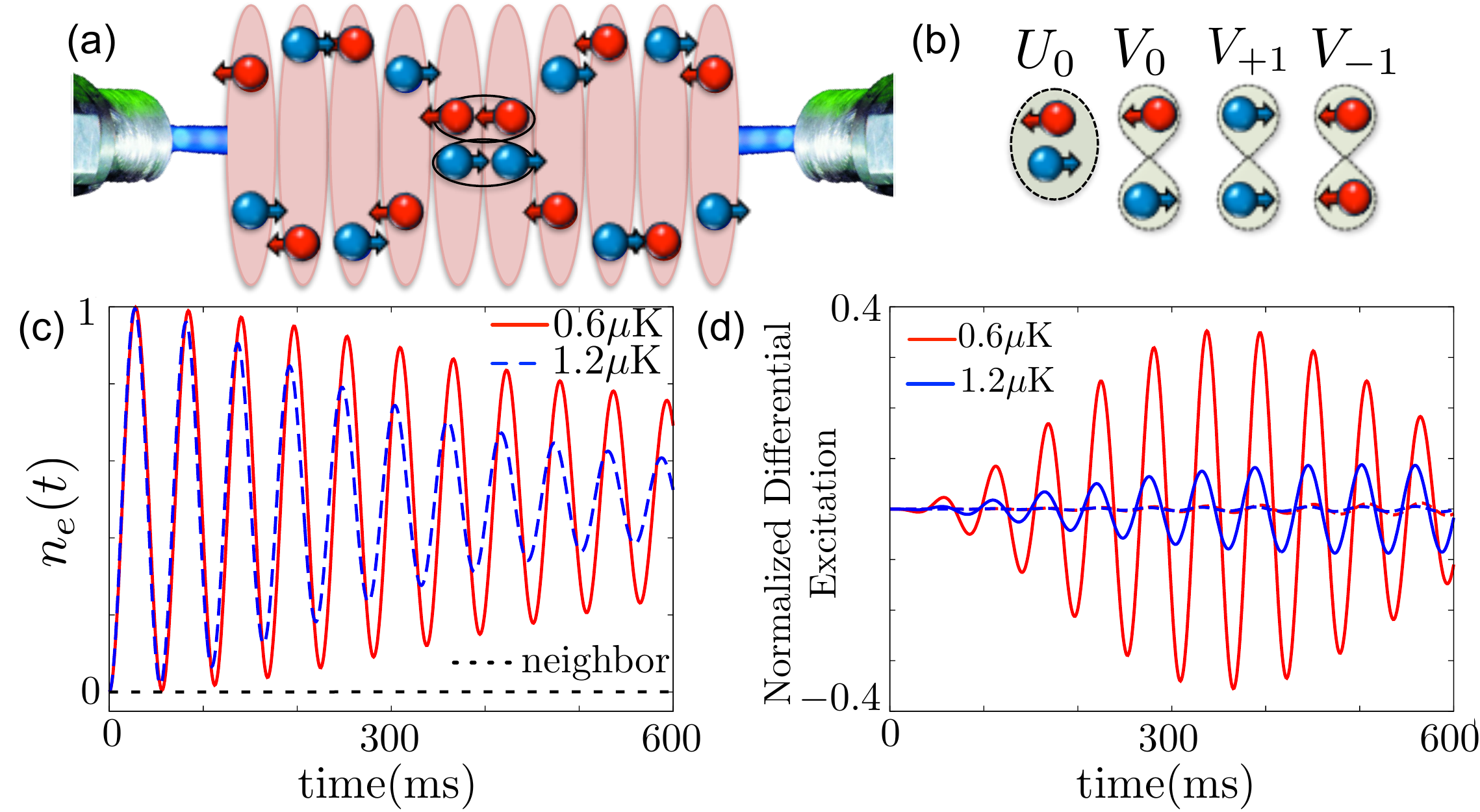}
\caption{ (a) An additional  counter-propagating  probing beam with a differential phase $\Upsilon$ generates a sliding superlattice potential, shown for $\phi=7\pi/6$, corresponding to an $^{87}$Sr OLC.  For weak tunneling $J\ll\Omega$ transport is energetically suppressed except at resonant defect points (circled).   (b) Two-particle interaction sectors classified by the total polarization $M_x$ and spatial symmetry of the dressed states.  An oval (figure-eight) denotes a symmetric (antisymmetric) spatial wavefunction.  (c) Dynamics of a single particle at the tunneling resonance for two temperatures (red solid and blue dashed) and an off-resonant site (black dotted) for $J_{\mathbf{0}}=8$Hz, $\Omega_{\mathbf{0}}=1$kHz, $\phi=7\pi/6$.  (d) Normalized differential excitation extracted as explained in the text with the interaction parameters of Ref.~\cite{SUN2}.  The $M_x=0$ components (solid lines) involve the $s$-wave sector, and so display a strong dependence of contrast on temperature, while the $M_{\pm1}$ sectors (dashed lines) experience only weaker single-particle thermal dephasing.}\label{fig:SSL}
\end{figure}

{\it Sliding Superlattice:}  We  consider the third case where the interrogation is done by  a pair of counter-propagating beams close to resonance with  the clock transition and with a global phase difference $\Upsilon(t)$ which can be controlled in time (see Fig. \ref{fig:SSL}(a)). The non-interacting  Hamiltonian, written using a Wannier orbital basis along the lattice direction, is
\begin{eqnarray}
&&\hat{H}_L^0=-\sum_{\mathbf{n},j,\alpha}\left(J_{\mathbf{n}}\left[\hat{a}^{\dagger}_{\alpha,\mathbf{n},j}\hat{a}_{\alpha,\mathbf{n},j+1}+\mathrm{H.c.}\right]-
\frac{\delta}{2}\alpha  \hat{n}_{\alpha\mathbf{n},j}\right)\notag\\
\label{eq:PumpingHami}&&-\sum_{\mathbf{n},j} \left(\Omega_{\mathbf{n}}\cos (\Upsilon(t)-j\phi)\left[\hat{a}^{\dagger}_{+,\mathbf{n},j}\hat{a}_{- ,\mathbf{n},j}+\mathrm{H.c.}\right]\right). \end{eqnarray}
As $\Upsilon(t)$ is changed, the clock laser standing wave ``slides" with respect to the optical lattice. This allows for the minimum realization of a topological pump when $\Upsilon(t)$ is adiabatically varied from $0\to2\pi$~\cite{Thouless,TroyerPump,KwekPump,MuellerPump,CooperRey}.  In the weak tunneling limit the quantized nature of particle transport can be directly linked to spatially isolated tunneling resonances ~\cite{CooperRey}.
We now show how those resonances can be spectroscopically measured in OLCs.



Let us  first consider the case  $J_{\mathbf{n}}=\delta=0$ and set  $\phi=7 \pi/6$,  relevant for the ${}^{87}$Sr system. We   write $\Upsilon= (  2 \pi s  +\Delta \Upsilon)/12$, with $0\leq \Delta \Upsilon<2\pi $ and  $s$ an integer.    Under these conditions, the localized  dressed eigenstates are spin-polarized along $\pm x$ alternating between neighboring sites except for ``defects''  at $j_d(r)=6 r+3 + s$ ($r$ an integer)  when  $\cos (\Upsilon-\Delta\Upsilon/12 -j_d\phi)=0$ and  the ground states at $j_d(r)$ and $j_d(r)+1$ point along the same direction.  Since tunneling preserves polarization, it is  suppressed when $J_{\mathbf{n}}\ll \Omega_{\mathbf{n}}$ due to the energy offset $\sim\Omega$ between neighboring sites.
The one exception is the case  $\Delta \Upsilon=\pi $ where the defect pair $j_d(r)$  and $j_d(r)+1$ is resonantly tunnel-coupled. Quantized transport occurs when $\Delta \Upsilon$ is slowly varied across the resonance. Instead of adiabatic transport we propose to  spectroscopically resolve the resonance   using a Ramsey-type protocol. Here, atoms initially prepared in $g$ at $|\delta|\gg \Omega_{\mathbf{0}}$ are adiabatically transferred to the ground dressed state  by slowly turning off $\delta$ at $ \Upsilon= (  2 \pi s  +\Delta \Upsilon_p)/12 $ with $\Delta \Upsilon_p\neq \pi$. Then  $ \Upsilon$  is quenched so resonant tunneling is allowed, $s\to s+1$ and $\Delta \Upsilon=\pi$, and the system evolves for a time $\tau$. Following this evolution tunneling is turned off and the phase switched back, $s+1\to s$ and $\Delta \Upsilon \to \Delta \Upsilon_p$. Those atoms which have tunneled  at the resonant sites are now in an excited state of the local dressed basis.  These excitations can be measured by adiabatically converting the dressed excitations to  ``bare''  $e$  excitations by adiabatically ramping on  $\delta\gg \Omega_{\mathbf{0}} $; this leads to a measurable excited state population $n_e(t)$.  In Fig. \ref{fig:SSL}(c) we show these tunneling resonances are clearly observable even at finite temperature.
Moreover, since resonances are spatially well-separated (every six sites), they can be resolved with low-resolution imaging.

Interactions modify the transport dynamics in our  protocol.  The sliding superlattice simplifies the treatment of interactions by isolating resonant site pairs $\{j_d,j_{d}+1\}$. We consider the case where at most  two atoms occupy the resonant sites, a condition which can be achieved by decreasing the atomic density with a large-volume dipole trap~\cite{Bloom}.   The two particle  states  can be classified in terms of  the atoms' spin polarization along $x$ in  four sectors: three symmetric ones (triplets) with total $x$ polarization  $M_x=+1,-1,0$ and a $M_x=0$ singlet. Within each sector, the dressed states' interaction parameters are given by~\cite{Supp}  $V_{+1}=V_{-1}=(V^{gg}+V^{ee}+2V^{eg})/4$, $V_{0}=(V^{gg}+V^{ee})/2$ and $U_{0}=U^{eg}$ (see Fig. \ref{fig:SSL}(b)). Since the $p$-wave  parameters are not SU$(2)$ symmetric, i.e. $b_{ee}\neq b_{gg}\neq b_{eg}$, the triplet sectors are coupled. However, in the weakly interacting limit  $V^{\alpha \beta}\ll \Omega$, the triplets are separated by energy gaps $\sim\Omega$ and coupling between the sectors can be neglected. The singlet sector is always decoupled from the triplets.

Within each sector interactions modify the dynamics making it sensitive to  temperature and density. The modifications  can be extracted by performing measurements of  the atom number-normalized excitation fraction for different densities and then differentiating the high-density and low-density results~\cite{Supp}. This  procedure removes the single-particle contribution and is particularly suitable  for characterizing the role of interactions in clock experiments~\cite{Martin_2013,SUN2}. The normalized differential excitation is shown in Fig.~\ref{fig:SSL}(d).  For the adiabatic dressed state preparation, all four manifolds, and hence all interaction parameters $V_{\pm,0},U_0$, contribute to the dynamics.  A filtering protocol that uses the $ee$ losses can be used to separate the dynamics of the various sectors. For example, by transferring all atoms to the $e$ state and holding before the adiabatic ground state preparation, the doubly occupied $M_x=\pm 1$ triplet sectors will be removed and only the $M_x=0$ singlet and triplet remain and contribute (here the ground dressed  states have one atom at $j_d$ and $j_d+1$). As shown in Fig.~\ref{fig:SSL} the dynamics of the  $V_{\pm,0}$ and $U_0$ sectors can be distinguished by the different scaling of the $p$ and $s$-wave interaction parameters with temperature $T_R$, $V_{\pm,0}\left(T_R\right)\sim \mathrm{const.}$ and $U_0\left(T_R\right)\sim T_R^{-1}$~\cite{Rey_Annals} (See Fig.\ref{fig:SSL}(d)).  By comparing these dynamics to that without the holding time, information about the $M_x=\pm1$ dynamics  can be extracted. In general $s$-wave interactions dominate the normalized differential contrast, with $p$-wave  contributions relevant only at hotter temperatures.

{ \it Summary } We have described approaches to implement and probe SOC in OLCs. We showed the proposed protocols  provide clean signatures of SOC in interacting many-body systems. While they work at the temperatures achievable in current OLCs,  their sensitivity  and applicability are expected to significantly improve  when operating the clock in the quantum degenerate regime. Moreover, if   nuclear spins are included, they can open   a window for the investigation of  SOC  with   SU(N)-symmetric  collisions~\cite{Cazalilla_rev}.

{\it Acknowledgments} We thank Ed Marti and James Thompson for comments on the manuscript, and Leonid Isaev and the JILA Sr clock team for discussions.  This work was supported by the NSF (PIF-1211914 and
  PFC-1125844), AFOSR, AFOSR-MURI, NIST and ARO, the  EPSRC Grant EP/K030094/1,  and by
  the JILA Visiting Fellows Program. MLW thanks the NRC postdoctoral fellowship program for support.  AK was supported by the Department of Defense through the NDSEG program.

\clearpage

\onecolumngrid

\renewcommand{\theequation}{S\arabic{equation}}
\setcounter{equation}{0}

\hrulefill
\begin{center}
\Large\textbf{Supplemental Material for ``Exploring synthetic spin-orbit coupling in a thermal optical lattice clock''}
\end{center}

\emph{Interaction Hamiltonian with the spin model assumption }
 The Hamiltonian governing fermionic AEAs in an optical lattice clock may be written:
\begin{align}
\label{eq:H}\hat{H}&=\hat{H}_0+\hat{H}_I+\hat{H}_L\, ,\\
 \hat{H}_0&=\sum_{\alpha} \int d\mathbf{r}\hat{\psi}_{\alpha}\left(\mathbf{r}\right)\left[-\frac{\hbar^2}{2m}\nabla^2+V_{\mathrm{ext}}\left(\mathbf{r}\right)\right]\hat{\psi}_{\alpha}\left(\mathbf{r}\right) -\hbar \delta\int d\mathbf{r}\hat{\psi}_{e}^{\dagger}\left(\mathbf{r}\right)\hat{\psi}_{e}\left(\mathbf{r}\right)\, ,\\
\label{eq:HI} \hat{H}_I&=\frac{4\pi \hbar^2a_{eg}^{-}}{m}\int d\mathbf{r}\hat{\psi}_{e}^{\dagger}\left(\mathbf{r}\right)\hat{\psi}_{e}\left(\mathbf{r}\right)\hat{\psi}_{g}^{\dagger}\left(\mathbf{r}\right)\hat{\psi}_{g}\left(\mathbf{r}\right)+\sum_{\alpha\beta}\frac{3\pi\hbar^2b^3_{\alpha\beta}}{m}\int d\mathbf{r}W\left[\hat{\psi}^{\dagger}_{\alpha}(\mathbf{r}),\hat{\psi}^{\dagger}_{\beta}(\mathbf{r})\right]\left(W\left[\hat{\psi}^{\dagger}_{\alpha}(\mathbf{r}),\hat{\psi}^{\dagger}_{\beta}(\mathbf{r})\right]\right)^{\dagger}\, ,\\
\hat{H}_L&=-\frac{\hbar \Omega}{2}\int d\mathbf{r} \left[\hat{\psi}_{e}^{\dagger}\left(\mathbf{r}\right)e^{i2\pi Z/\lambda}\hat{\psi}_{g}\left(\mathbf{r}\right)+\mathrm{H.c.}\right]
\end{align}
where $m$ is the atomic mass, $\hat{\psi}_{\alpha}\left(\mathbf{r}\right)$ is a fermionic field operator for state $\alpha\in \left\{g,e\right\}$, $W\left[\hat{A}\left(\mathbf{r}\right),\hat{B}\left(\mathbf{r}\right)\right]=(\nabla \hat{A}\left(\mathbf{r}\right))\hat{B}\left(\mathbf{r}\right)-\hat{A}\left(\mathbf{r}\right)(\nabla \hat{B}\left(\mathbf{r}\right))$ is the Wronskian, and $\delta=\omega_l-\omega_0$ is the difference between the clock laser frequency $\omega_l$ and atomic frequency $\omega_0$ in the rotating frame of the laser.  The Hamiltonian $\hat{H}_0$ contains the kinetic energy and trapping potential $V_{\mathrm{ext}}\left(\mathbf{r}\right)$, $\hat{H}_I$ contains the effects of $s$-wave interactions with scattering length $a_{eg}^-$ and $p$-wave interactions with interaction volumes $b_{\alpha\beta}^3$ between nuclear spin-polarized AEAs, and $\hat{H}_L$ describes the coupling of internal levels to the clock laser.  We assume that the optical lattice is deep enough that we can neglect interactions occurring between different lattice sites, and so consider only interactions between particles occupying the same lattice site.  Further, we enact the spin model approximation~\cite{Rey_Annals}, which neglects all interactions which do not preserve the single-particle transverse mode occupations during a two-body collision.  The spin model Hamiltonian hence keeps only direct terms, in which the two colliding particles remain in their same transverse motional quantum states, and exchange processes where the transverse quantum numbers of the two colliding particles are exchanged.  The spin model Hamiltonian is valid when interactions are smaller than the spacing between transverse modes, and also when anharmonicity of the potential prevents collisional exchange of mode energy between transverse dimensions.

To derive a Hubbard model amenable for direct calculation, we expand the field operators in a basis of single-particle eigenfunctions.  In order to facilitate thermal averages, we take this set of functions to be the eigenfunctions of $\hat{H}_0$, $\psi_{q,n}\left(\mathbf{r}\right)$, which are indexed in terms of a quasimomentum $q$ and a transverse mode index $\mathbf{n}$.  Enacting this expansion with the approximations of the last paragraph, we find
\begin{align}
\nonumber \hat{H}_I&=\frac{1}{4L}\sum_{\alpha \beta}\sum_{\left\{\mathbf{n}_1,\mathbf{n}_2\right\}}\sum_{qq'\Delta q}\Big[\left(2-\delta_{\mathbf{n}_1\mathbf{n}_2}\right)\left(U^{\alpha \beta}_{\left\{\mathbf{n}_1,\mathbf{n}_2\right\}}+V^{\alpha\beta}_{\left\{\mathbf{n}_1,\mathbf{n}_2\right\}}\right)\hat{a}^{\dagger}_{\alpha,q+\Delta q,\mathbf{n}_1}\hat{a}^{\dagger}_{\beta,q'-\Delta q,\mathbf{n}_2}\hat{a}_{\beta,q',\mathbf{n}_2}\hat{a}_{\alpha,q,\mathbf{n}_1}\\
\label{eq:SpinModelInt}&+2\left(1-\delta_{\mathbf{n}_1\mathbf{n}_2}\right)\left(V^{\alpha\beta}_{\left\{\mathbf{n}_1,\mathbf{n}_2\right\}}-U^{\alpha \beta}_{\left\{\mathbf{n}_1,\mathbf{n}_2\right\}}\right)\hat{a}^{\dagger}_{\beta,q'-\Delta q,\mathbf{n}_1}\hat{a}^{\dagger}_{\alpha,q+\Delta q,\mathbf{n}_2}\hat{a}_{\beta,q',\mathbf{n}_2}\hat{a}_{\alpha,q,\mathbf{n}_1}\Big]\, .
\end{align}
Here, $L$ is the number of lattice sites, the sum over $\left\{\mathbf{n}_1,\mathbf{n}_2\right\}$ means the sum over distinct, ordered pairs of transverse modes, $q$, $q'$, and $\Delta q$ are quasimomenta in the first Brillouin zone (BZ), greek letters denote internal electronic states $\alpha,\beta\in\left\{g,e\right\}$, and the interaction matrix elements may be written as
\begin{align}
U^{\alpha \beta}_{\left\{\mathbf{n}_1,\mathbf{n}_2\right\}}&=\left(1-\delta_{\alpha\beta}\right)\frac{8\pi \hbar^2}{m}a_s S_{\mathbf{n}_1\mathbf{n}_2\mathbf{n}_2\mathbf{n}_1}\, ,\;\;V^{\alpha\beta}_{\left\{\mathbf{n}_1,\mathbf{n}_2\right\}}=\frac{12\pi\hbar^2}{m}b^3_{\alpha\beta}P_{\mathbf{n}_1\mathbf{n}_2\mathbf{n}_2\mathbf{n}_1}
\end{align}
by defining the integrals
\begin{align}
S_{\mathbf{n}_1'\mathbf{n}_2'\mathbf{n}_2\mathbf{n}_1}&=\int d\mathbf{r} \psi_{i,\mathbf{n}_1'}^{\star}\left(\mathbf{r}\right) \psi_{i,\mathbf{n}_2'}^{\star}\left(\mathbf{r}\right) \psi_{i,\mathbf{n}_2}\left(\mathbf{r}\right)\psi_{i,\mathbf{n}_1}\left(\mathbf{r}\right)\, ,\;\;P_{\mathbf{n}_1'\mathbf{n}_2'\mathbf{n}_2\mathbf{n}_1}=\int d\mathbf{r}W\left[\psi^{\star}_{i,\mathbf{n}_1'}\left(\mathbf{r}\right),\psi^{\star}_{i,\mathbf{n}_2'}\left(\mathbf{r}\right)\right] W\left[\psi_{i,\mathbf{n}_1}\left(\mathbf{r}\right),\psi_{i,\mathbf{n}_2}\left(\mathbf{r}\right)\right]\, .
\end{align}
In the latter two expressions, the wavefunctions are localized Wannier-type functions obtained from the quasimomentum-indexed eigensolutions (Bloch functions) of the full 3D potential as $\psi_{j,\mathbf{n}}\left(\mathbf{r}\right)=\frac{1}{\sqrt{L}}\sum_{q\in \mathrm{BZ}}e^{-i q r_j}\psi_{q,\mathbf{n}}\left(\mathbf{r}\right)$~\cite{Wall_Hazzard_15}.  We can also define the mode-dependent tunneling and Rabi frequency in terms of these functions as
\begin{align}
J_{\mathbf{n}}&=-\int d\mathbf{r} \psi_{j,\mathbf{n}}\left(\mathbf{r}\right)\left[-\frac{\hbar^2}{2m}\nabla^2+V_{\mathrm{ext}}\left(\mathbf{r}\right)\right] \psi_{j+1,\mathbf{n}}\left(\mathbf{r}\right)\, ,\\
\Omega_{\mathbf{n}}&=\Omega \int d\mathbf{r}\psi_{j,\mathbf{n}}\left(\mathbf{r}\right) e^{i2\pi Z/\lambda}\psi_{j,\mathbf{n}}\left(\mathbf{r}\right)\, .
\end{align}

\emph{First-order density shift in Ramsey spectroscopy}
In this section we compute the density shift for Ramsey spectroscopy to first order in interactions using an interaction picture perturbation series.  In particular, we write the propagator during the dark time evolution in terms of a truncated Dyson series $e^{-i\hat{H}t}=e^{-i\hat{H}_0t}-i \int_{0}^{t}dt' e^{-i\hat{H}_0\left(t-t'\right)}\hat{H}_1 e^{-i \hat{H}_0t'}.$  At the end of the dark time, a spin-rotation pulse of area $\theta_2$ is applied, and then the total $\hat{S}^z$ is measured.  These two operations can be combined as the measurement of the operator $\hat{S}^z_{\theta}=\sum_{nq}\left[\cos\theta\hat{S}^z_{q\mathbf{n}}-\sin\theta \hat{S}^y_{q\mathbf{n}}\right]$.  Hence, the first-order result of the Ramsey sequence is
\begin{align}
&\langle \hat{S}^z\left(\tau\right)\rangle=\langle \psi\left(\theta_1\right)|e^{i\hat{H}_0 \tau}\hat{S}^z_{\theta_2}e^{-i\hat{H}_0 \tau}|\psi\left(\theta_1\right)\rangle +2\mathcal{I}\left[\langle \psi\left(\theta_1\right)|e^{i\hat{H}_0 \tau}\hat{S}^z_{\theta_2}\int_0^{\tau} dt e^{-i \hat{H}_0\left(\tau-t\right)}\hat{H}_I e^{-i\hat{H}_0 t}|\psi\left(\theta_1\right)\rangle \right]\, ,
\end{align}
where $|\psi\left(\theta_1\right)\rangle$ is the state resulting from applying the first Ramsey pulse to an initial state $|\tilde{i}\rangle$.  We will parameterize the initial state in terms of products of single-particle eigenstates $|i\rangle=\prod_{i=1}^{N} \hat{a}^{\dagger}_{g q_in_i}|\mathrm{vac.}\rangle$, where $|\mathrm{vac}.\rangle$ is the vacuum, $N$ the number of particles, and $\left\{q_i,\mathbf{n}_i\right\}$ the initial distinct set of populated quasimomenta and transverse modes, so that
\begin{align}
|\psi\left(\theta_1\right)\rangle&=\prod_{j=1}^{N} \left(\cos\frac{\theta_1}{2}\hat{a}^{\dagger}_{g,q_j,\mathbf{n}_j}+i \sin\frac{\theta_1}{2}\hat{a}^{\dagger}_{e,q_j,\mathbf{n}_j}\right)\, .
\end{align}
From this, we find that the non-interacting dynamics are given as
\begin{align}
&\langle \psi\left(\theta_1\right)|e^{i\hat{H}_0 \tau}\hat{S}^z_{\theta_2}e^{-i\hat{H}_0 \tau}|\psi\left(\theta_1\right)\rangle=\sum_{i=1}^{N}\left[-\frac{\cos\theta_1\cos\theta_2}{2}+\frac{\sin\theta_1\sin\theta_2}{2}\cos\left(\delta \tau-\Delta E_{\mathbf{n}_i}\left(q_i,\phi\right)\tau\right)\right]\, ,
\end{align}
where $E_{\mathbf{n}}\left(q\right)$ is the single-particle dispersion and
\begin{align}
\label{eq:DeltaE}\Delta E_{\mathbf{n}}\left(q,\phi\right)\equiv E_{\mathbf{n}}\left(q+\phi\right)-E_{\mathbf{n}}\left(q\right)\, ,
\end{align}
is the difference in single-particle dispersion of the $e$ and $g$ states.

Writing the non-interacting result as $\langle \hat{S}^z\left(\tau\right)\rangle =A\left(\tau\right)\cos\left(\delta \tau\right)+B\left(\tau\right)\sin\left(\delta \tau\right)+\mathrm{const.}$ we can extract the density shift as $\Delta \nu=\frac{1}{2\pi \tau}\mathrm{arctan}\left(\frac{B\left(\tau\right)}{A\left(\tau\right)}\right)$ and the normalized contrast decay as $\mathcal{C}\left(\tau\right)=\frac{\sqrt{A^2\left(\tau\right)+B^2\left(\tau\right)}}{\sqrt{A^2\left(0\right)+B^2\left(0\right)}}\, .$  For a single particle with momentum $q$ and transverse mode $n$, the density shift is $\Delta E_{\mathbf{n}}\left(q,\phi\right)/2\pi$.  Assuming short times compared to the tunneling bandwidth, the density shifts for all particles add as $\Delta \nu=\sum_{i=1}^{N}\Delta E_{\mathbf{n}_i}\left(q_i,\phi\right)/2\pi N+\mathcal{O}\left(\tau^2\right)$.  The contrast decay at short times is given as $\mathcal{C}\left(\tau\right)=1-\frac{1}{2 N^2}\left(\sum_{i=1}^{N}\Delta E_{\mathbf{n}_i}\left(q_i,\phi\right)\right)^2\tau^2+\frac{1}{2 N}\sum_{i=1}^{N} \Delta E_{\mathbf{n}_i}^2\left(q_i,\phi\right) \tau^2 +\mathcal{O}\left(\tau^4\right)$.  This single-particle contribution will dominate over the $\sim U^2\tau^2$ decay of the contrast due to interactions at short times and in the regime of weak interactions compared to the tunneling.

The contribution to the dynamics at the lowest order in interactions is given as
\begin{align}
\label{eq:FullRamsey}&2\mathcal{I}\left[\langle \psi\left(\theta_1\right)|e^{i\hat{H}_0 \tau}\hat{S}^z_{\theta_2}\int_0^{\tau}dt e^{-i \hat{H}_0\left(\tau-t\right)}\hat{H}_I e^{-i\hat{H}_0 t}|\psi\left(\theta_1\right)\rangle\right]\\
\nonumber &=\frac{\tau}{L}\sum_{\left\{p_1,p_2\right\}}\Big\{ \sin\theta_1\sin\theta_2\sin\left[\left(\delta-\frac{\dE{p_1}{p_1}+\dE{p_2}{p_2}}{2}\right)\tau\right]\\
\nonumber&\times \Big[\left(C_{\left\{\mathbf{n}_{p_1}\mathbf{n}_{p_2}\right\}}-\chi_{\left\{\mathbf{n}_{p_1}\mathbf{n}_{p_2}\right\}}\cos\theta_1\right)\cos\left(\frac{\dE{p_1}{p_1}-\dE{p_2}{p_2}}{2}\tau\right)\\
\nonumber &-\frac{\cos\theta_1}{2}\left(2-\delta_{\mathbf{n}_{p_1}\mathbf{n}_{p_2}}\right)\zeta_{\left\{\mathbf{n}_{p_1}\mathbf{n}_{p_2}\right\}}\Big[\cos\left(\frac{\dE{p_1}{p_1}-\dE{p_2}{p_2}}{2}\tau\right)-\sinc\left(\frac{\dE{p_1}{p_1}-\dE{p_2}{p_2}}{2}\tau\right)\Big]\Big]
\Big\}
\end{align}
where the mode-dependent spin model parameters are
\begin{align}
C_{\left\{\mathbf{n}_1\mathbf{n}_2\right\}}&=\frac{\V{ee}{1}{2}-\V{gg}{1}{2}}{2}\\
\chi_{\left\{\mathbf{n}_1\mathbf{n}_2\right\}}&=\frac{\V{ee}{1}{2}+\V{gg}{1}{2}-2\V{ge}{1}{2}}{2}\\
\zeta_{\left\{\mathbf{n}_1\mathbf{n}_2\right\}}&=\frac{\V{eg}{1}{2}-\U{ge}{1}{2}}{2}\, .
\end{align}
Taking a series in the tunneling bandwidth, the zeroth order term for a given pair of particles is $\left(C-\chi\cos\theta_1\right)\sin\theta_1\sin\theta_2\sin\delta \tau$, which has been obtained in previous works~\cite{Martin_2013}.  The thermally averaged density shift in the high-temperature limit will be given in the next section.

\emph{High-temperature thermal average of density shift}
Here, we derive the high-temperature (compared to the tunneling bandwidth) thermal average of the first-order many-body dynamics derived in the last section.  First, we expand the above result to third order in time, finding
\begin{align}
\label{eq:FullThirdOrder}\langle \hat{S}^z\left(\tau\right)\rangle&=\frac{1}{2}\sum_{i=1}^{N}\left(\cos\left(\delta \tau\right)\sin\theta_1\sin\theta_2-\cos\theta_1\cos\theta_2\right)\\
\nonumber &+\sum_{\left\{p_1,p_2\right\}}\frac{\sin\theta_1\sin\theta_2}{L}\tau\left(C_{\left\{\mathbf{n}_{p_1},\mathbf{n}_{p_2}\right\}}-\chi_{\left\{\mathbf{n}_{p_1},\mathbf{n}_{p_2}\right\}} \cos\theta_1\right)\sin\delta \tau+\frac{1}{2}\sum_{i=1}^{N} \Delta E_{\mathbf{n}_i}\left(q_i,\phi\right)\sin\left(\delta \tau\right) \sin\theta_1\sin\theta_2 \tau\\
\nonumber &-\sum_{\left\{p_1,p_2\right\}}\frac{\sin\theta_1\sin\theta_2}{L}\frac{\tau^2}{2}\left(C_{\left\{\mathbf{n}_{p_1},\mathbf{n}_{p_2}\right\}}-\chi_{\left\{\mathbf{n}_{p_1},\mathbf{n}_{p_2}\right\}}\cos\theta_1\right)\left(\dE{p_1}{p_1}+\dE{p_2}{p_2}\right)\cos\delta \tau\\
\nonumber&-\frac{1}{4}\sum_{i=1}^{N} \Delta E_{\mathbf{n}_i}\left(q_i,\phi\right)^2\cos\left(\delta \tau\right) \sin\theta_1\sin\theta_2 \tau^2\\
\nonumber &-\sum_{\left\{p_1,p_2\right\}}\frac{\sin\theta_1\sin\theta_2}{L}\frac{\tau^3}{24}\Big[ 6\left(C_{\left\{\mathbf{n}_{p_1},\mathbf{n}_{p_2}\right\}}-\chi_{\left\{\mathbf{n}_{p_1},\mathbf{n}_{p_2}\right\}}\cos\theta_1\right)\left(\dE{p_1}{p_1}^2+\dE{p_2}{p_2}^2\right)\\
\nonumber &-\left(2-\delta_{\mathbf{n}_{p_1}\mathbf{n}_{p_2}}\right)\zeta_{\left\{\mathbf{n}_{p_1},\mathbf{n}_{p_2}\right\}}\left(\dE{p_1}{p_1}-\dE{p_2}{p_2}\right)^2\cos\theta_1\sin\delta \tau\Big]\\
&\nonumber-\frac{1}{12}\sum_{i=1}^{N} \Delta E_{\mathbf{n}_i}\left(q_i,\phi\right)^3\sin\left(\delta \tau\right) \sin\theta_1\sin\theta_2 \tau^3
\Big\}\, .
\end{align}
To perform thermal averages, we now enact the tight-binding approximation, in which $E_{\mathbf{n}}\left(q\right)=\bar{E}_{\mathbf{n}}-2 J_{\mathbf{n}}\cos\left(q\right)$, and compute the partition function as
\begin{align}
Z&=\sum_{nq}e^{-\beta E_{\mathbf{n}}\left(q\right)}\approx \sum_{\mathbf{n}} e^{-\beta \bar{E}_{\mathbf{n}}}  \int_{-1}^1 dq e^{2\beta J_{\mathbf{n}} \cos\left(\pi q\right)}\\
&=2\sum_{\mathbf{n}} e^{-\beta \bar{E}_{\mathbf{n}}} I_0\left(2\beta J_{\mathbf{n}}\right)=2\sum_{\mathbf{n}}e^{-\beta \bar{E}_{\mathbf{n}}}\left[1+\mathcal{O}\left(\beta^2 J_{\mathbf{n}}^2\right)\right]\, ,
\end{align}
where $I_{\nu}\left(x\right)$ is the modified Bessel function of order $\nu$ and $\beta$ the inverse radial temperature.  Hence, for temperatures sufficiently high compared to the bandwidth $J_{\mathbf{n}}$, the partition function is just a constant times the partition function of the transverse modes.  Now,
\begin{align}
\label{eq:JinfTemp}\langle \Delta E_n\left(q,\phi\right)\rangle_{T_R}&=\frac{1}{Z}\sum_{\mathbf{n}}e^{-\beta \bar{E}_{\mathbf{n}}} 4J_{\mathbf{n}}\left(1-\cos\phi\right)I_1\left(2\beta J_{\mathbf{n}}\right)\\
&=\frac{1}{Z_{\mathbf{n}}}\sum_{\mathbf{n}} e^{-\beta \bar{E}_{\mathbf{n}}} 2 \beta J_{\mathbf{n}}\left(1-\cos\phi\right)+\mathcal{O}\left(\beta^2J_{\mathbf{n}}^2\right)\, ,
\end{align}
where $Z_{\mathbf{n}}=\sum_{\mathbf{n}}e^{-\beta \bar{E}_{\mathbf{n}}}$ is the partition function of the transverse modes.  This result gives that terms which are linear in $\Delta E_{\mathbf{n}}\left(q,\phi\right)$ for a specific $q$, including terms like $\dE{1}{1}\dE{2}{2}$, vanish at least as fast as $\mathcal{O}\left(\beta\right)$ at high temperatures.  In contrast, we find
\begin{align}
\nonumber & \langle \Delta E_{\mathbf{n}}^2\left(q,\phi\right)\rangle_{T_R}=\frac{1}{Z}\sum_{\mathbf{n}}e^{-\beta \bar{E}_{\mathbf{n}}}16J_{\mathbf{n}}^2\Big[\frac{1}{\beta J_{\mathbf{n}}}I_1\left(2\beta J_{\mathbf{n}}\right)+I_2\left(2\beta J_{\mathbf{n}}\right)\left(1-\cos\phi\right)\Big]\sin^2\frac{\phi}{2}\\
&=\frac{1}{Z_{\mathbf{n}}}\sum_{\mathbf{n}}e^{-\beta \bar{E}_{\mathbf{n}}}4J_{\mathbf{n}}^2\left(1-\cos\phi\right)+\mathcal{O}\left(\beta^2 J_{\mathbf{n}}^2\right)\, ,
\end{align}
and so the thermal average of $\Delta E^2_{\mathbf{n}}\left(q,\phi\right)$ is a constant to lowest order in a high-temperature expansion.

Using these results, we find that the leading order approximation in a high-temperature series expansion is
\begin{align}
\nonumber \langle \langle \hat{S}^z_{\theta_2}\rangle\rangle_{T_R}&=\frac{N}{2}\left(\sin\theta_1\sin\theta_2\cos\left(\delta \tau\right)-\cos\theta_1\cos\theta_2\right)+\frac{\tau N\left(N-1\right)}{2L}\sin\theta_1\sin\theta_2\sin\left(\delta \tau\right)\left( \langle{C}\rangle_{T_R}- \langle{\chi}\rangle_{T_R} \cos\theta_1\right)\\
&-2N\langle{J^2}\rangle_{T_R}\sin\theta_1\sin\theta_2 \cos\left(\delta \tau\right)\sin^2\frac{\phi}{2}\tau^2-\frac{\langle{J^2}\rangle_{T_R} \tau^3N\left(N-1\right)}{6}\sin^2\frac{\phi}{2}\left[6\left(\langle{C}\rangle_{T_R}-\langle{\chi}\rangle_{T_R}\cos\theta_1\right)-2\langle{\zeta}\rangle_{T_R} \cos\theta_1\right]\, ,
\end{align}
where we have neglected interactions between pairs of particles with the same transverse mode for simplicity, $\langle \bullet\rangle_{T_R}$ denotes a thermal average with respect to the transverse modes, and we have assumed that, e.g.~$\langle{J^2 C}\rangle_{T_R}\approx \langle{J^2}\rangle_{T_R}\,\langle{C}\rangle_{T_R}$, which is valid when the tunneling is only weakly temperature dependent.  From this, we find the density shift
\begin{align}
\Delta \nu &=\Delta \nu_0\left[1+\frac{4}{3}\langle{J^2}\rangle_{T_R}\tau^2\sin^2\frac{\phi}{2}\frac{\langle{\zeta}\rangle_{T_R} \cos\theta_1}{\langle{C}\rangle_{T_R}-\langle{\chi}\rangle_{T_R}\cos\theta_1}\right]\, ,
\end{align}
where $\Delta \nu_0$ is the density shift in the absence of tunneling.

\emph{Lineshape and extraction of $\theta_{\mathbf{n}q}$ from momentum-resolved Rabi spectroscopy}

\begin{figure}
\centering
\includegraphics[width=0.5\columnwidth]{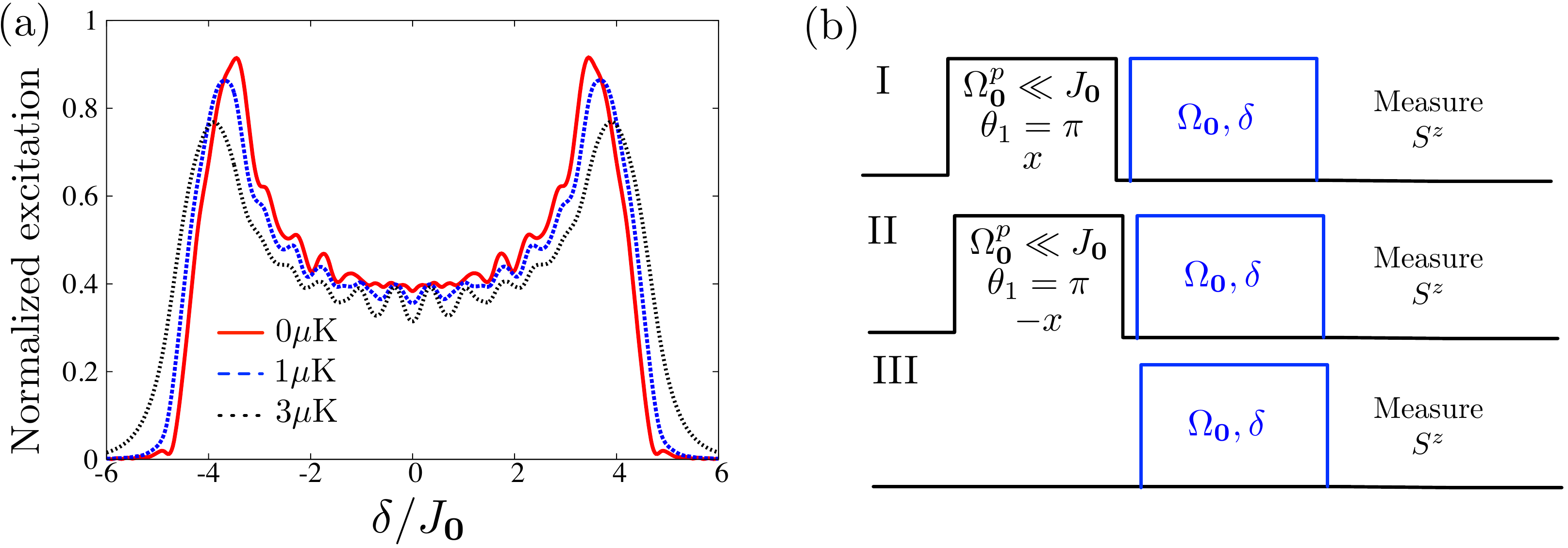}
\caption{(Color online)  (a)  Thermally averaged Rabi lineshape fully accounting for trap non-separability.  SOC-induced peaks are visible even at $T=3\mu$K.  (b) Three pulse sequences used to extract $\theta_{\mathbf{0}q}$ in Fig.~(2) of the main text.
}
\label{fig:SF1}
\end{figure}

As described in the main text, SOC introduces substructure in the Rabi lineshape within a window of width $8J_{\mathbf{n}}\left|\sin\frac{\phi}{2}\right|$ of the carrier frequency.  At finite temperature, many transverse modes are populated and hence the dependence of $\Omega_{\mathbf{n}}$ and $J_{\mathbf{n}}$ on $\mathbf{n}$ could  broaden the line and destroy this substructure. However, a direct simulation of the Rabi lineshape using the potential Eq.~\eqref{eq:LattPot} demonstrates that the features of the ideal, zero-temperature lineshape are captured up to a temperature of $3\mu$K as shown in Fig.~\ref{fig:SF1}(a). Thanks to the clock's sub-Hz resolution it will be  possible to resolve the lineshape for a wide range of lattice parameters.

We now turn to the extraction of the chiral Bloch vector angle $\theta_{\mathbf{n}q}$ with Rabi spectroscopy.  The eigenstates of Eq.~\eqref{eq:Bsprs} may be written as
\begin{align}
|\psi_{n,q,-}\rangle&=\cos\frac{\theta_{nq}}{2}|gqn\rangle+\sin\frac{\theta_{nq}}{2}|eqn\rangle\\
|\psi_{n,q,+}\rangle&=-\sin\frac{\theta_{nq}}{2}|gqn\rangle+\cos\frac{\theta_{nq}}{2}|eqn\rangle\, ,
\end{align}
with energies
\begin{align}
E_{n,q,\pm}&=\frac{E_{n}\left(q\right)+E_{n}\left(q+\phi\right)+\delta}{2}\pm \left|\mathbf{B}_{nq}\right|\, ,
\end{align}
and magnetization
\begin{align}
\langle \psi_{n,q,\pm}|\hat{S}^z_{n,q}|\psi_{n,q,\pm}\rangle&=\frac{\pm \cos\theta_{nq}}{2}\, .
\end{align}
Using the above, the expected Rabi dynamics in the ideal case for an arbitrary initial state $\alpha |g\mathbf{n}q\rangle+\beta|e\mathbf{n}q\rangle$ is found to be
\begin{align}
\nonumber S^z\left(\alpha,\beta,t\right)&=\frac{1}{2}\left(\left|\beta\right|^2-\left|\alpha\right|^2\right)\left[\cos^2\theta_{nq}+\cos\left(2B_{nq} t\right)\sin^2\theta_{nq}\right]+\mathcal{R}\left(\alpha^{\star}\beta\right)\cos\theta_{nq}\sin\theta_{nq}\left(\cos\left(2B_{nq}t\right)-1\right)\\
&+\mathcal{I}\left(\alpha^{\star}\beta\right)\sin\theta_{nq}\sin\left(2B_{nq}t\right)\, .
\end{align}

The protocol extracting the chiral Bloch vector angle $\theta_{\mathbf{n}q}$ involves three sequences (Fig.~\ref{fig:SF1}(b)). All sequences start by preparing the atoms in $g$. In sequence I,  a narrow $\pi$-pulse about the $x$ axis  is applied at the detuning  $\delta^{\star}$ associated with the $q^{\star}_{\mathbf{0}}$ resonance. The  Rabi frequency of the pulse, $\Omega_{\mathbf{0}}^p$, should be weak enough to  guarantee that only  atoms within a narrow window centered  around  $q^{\star}_\mathbf{0}$  are transferred to $e$. Atoms with $q\ne q^{\star}_\mathbf{0}$  are off-resonant and remain in $g$.  Next the detuning and Rabi frequency are quenched to  the desired values  $\Omega_{\mathbf{0}}$ and $\delta$, and Rabi oscillations are recorded.  Sequence  II is identical, except that the initial pulse is about $-x$.  Sequence III uses no initial  pulse, but is otherwise the same as sequence I.  The dynamics in  I contains information about $\theta_{q^{\star}_\mathbf{0}}$, but this information will be buried in the signal of the $q\ne q^{\star}_\mathbf{0}$ atoms.  Subtracting the dynamics of I and III isolates the dynamics of $q^{\star}_\mathbf{0}$ atoms.  Sequence  II is required because due to the mode dependence of $\Omega_{\mathbf{n}}$, a $\pi$-pulse is not experienced by all atoms populated at the transverse modes at high temperature. The dynamics of the atoms with $ q^{\star}_\mathbf{0}$   that remain in $g$ is cancelled by taking the average between sequences I and II, which  effectively deals with only the atoms transferred to $e$. The dynamics for the sequences I, II, and III given in the main text are $S^z\left(\sqrt{1-f_{\mathbf{n}}^2},-if_{\mathbf{n}},t\right)$, $S^z\left(\sqrt{1-f_{\mathbf{n}}^2},if_{\mathbf{n}},t\right)$, and $S^z\left(1,0,t\right)$, respectively, where $f_{\mathbf{n}}\approx 1$ is the excitation fraction resulting from the initialization $\pi$-pulse.  This gives the subtracted signal
\begin{align}
\label{eq:Szdyn}\frac{S^z\left(\sqrt{1-f_{\mathbf{n}}^2},-if_{\mathbf{n}},t\right)+S^z\left(\sqrt{1-f_{\mathbf{n}}^2},if_{\mathbf{n}},t\right)}{2}-S^z\left(1,0,t\right)&=f_{\mathbf{n}}^2\left(\cos^2\theta_{\mathbf{n}q}+\cos\left(2|B_{\mathbf{n}q}|t\right)\sin^2\theta_{\mathbf{n}q}\right)\, .
\end{align}
For a single frequency $|B_{\mathbf{n}q}|$, and  $\theta_{\mathbf{n} q}$ can be inferred from these oscillations as $2\tan^2\theta_{\mathbf{n} q}=(\mbox{max}-\mbox{min})/\mbox{mean}$.  In the multiple-frequency case this is no longer a strict equality, but Fig.~\ref{fig:2}(b) shows that determining $\theta_{\mathbf{n}q}$ in this fashion from the thermal average of Eq.~\eqref{eq:Szdyn} is nevertheless robust.  In addition, the validity of this approach can be quantitatively estimated by the equality $(\mbox{max}+\mbox{min})/\mbox{mean}=2$, which employs the same assumptions.

\emph{Interacting dressed states with a sliding clock superlattice} Here, we consider the dynamics of two atoms in neighboring sites $j_d(r)$ and $j_d(r)+1$ which are tunnel-coupled for a sliding lattice clock phase $\Delta \Upsilon=\pi$.  We will take these two atoms to have different transverse mode indices $\mathbf{n}$ and $\mathbf{m}$.  Interactions in the $s$-wave channel occur between these atoms when they are in an antisymmetric electronic state $|\Lambda^-_{\mathrm{eg}}\rangle\equiv \left(|ge\rangle-|eg\rangle\right)/\sqrt{2}$ with a symmetric spatial wave function $|\Psi^+_{\mathbf{nm}}\rangle\equiv \left(|\mathbf{nm}\rangle_j+|\mathbf{mn}\rangle_j\right)$, with the subscript $j$ denoting the lattice site index.  Instead, $p$-wave interactions occur when the atoms are in a symmetric electronic state $|\Lambda^+_{\mathrm{eg}}\rangle\equiv \left(|ge\rangle+|eg\rangle\right)/\sqrt{2}$ and an antisymmetric spatial wave function $|\Psi^-_{\mathbf{nm}}\rangle\equiv \left(|\mathbf{nm}\rangle_j-|\mathbf{mn}\rangle_j\right)$.  For the purposes of describing the system dynamics at $\Delta\Upsilon=\pi$, where $|\rightarrow\rangle=\left(|g\rangle+|e\rangle\right)/\sqrt{2}$ is the single-well electronic ground state and $|\leftarrow\rangle=\left(|g\rangle-|e\rangle\right)/\sqrt{2}$ the excited state, it is useful to employ the states
\begin{align}
t,1&: |2^{++},0\rangle,|0,2^{++}\rangle, |\rightarrow_{\mathbf{n}},\rightarrow_{\mathbf{m}}\rangle, |\rightarrow_{\mathbf{m}},\rightarrow_{\mathbf{n}}\rangle\, ,\\
\nonumber t,-1&: |2^{--},0\rangle,|0,2^{--}\rangle, |\leftarrow_{\mathbf{n}},\leftarrow_{\mathbf{m}}\rangle, |\leftarrow_{\mathbf{m}},\leftarrow_{\mathbf{n}}\rangle\\
\nonumber t,0&: |2^{\pm},0\rangle,|0,2^{\pm}\rangle, \frac{|\rightarrow_{\mathbf{n}},\leftarrow_{\mathbf{m}}\rangle-|\rightarrow_{\mathbf{m}},\leftarrow_{\mathbf{n}}\rangle}{\sqrt{2}},\frac{|\leftarrow_{\mathbf{n}},\rightarrow_{\mathbf{m}}\rangle-|\leftarrow_{\mathbf{m}},\rightarrow_{\mathbf{n}}\rangle}{\sqrt{2}}\\
\nonumber s,0&: |2^{\mp},0\rangle,|0,2^{\mp}\rangle, \frac{|\rightarrow_{\mathbf{n}},\leftarrow_{\mathbf{m}}\rangle+|\rightarrow_{\mathbf{m}},\leftarrow_{\mathbf{n}}\rangle}{\sqrt{2}},\frac{|\leftarrow_{\mathbf{n}},\rightarrow_{\mathbf{m}}\rangle+|\leftarrow_{\mathbf{m}},\rightarrow_{\mathbf{n}}\rangle}{\sqrt{2}}\, ,
\end{align}
where
\begin{align}
|2^{++}\rangle&=|\rightarrow\rightarrow\rangle|\Psi^-_{\mathbf{nm}}\rangle\, ,\\
|2^{--}\rangle&=|\leftarrow\leftarrow\rangle|\Psi^-_{\mathbf{nm}}\rangle\, ,\\
|2^{\pm}\rangle&=\frac{|\rightarrow\leftarrow\rangle+|\leftarrow\rightarrow\rangle}{\sqrt{2}}|\Psi^-_{\mathbf{nm}}\rangle\, ,\\
|2^{\mp}\rangle&=\frac{|\rightarrow\leftarrow\rangle-|\leftarrow\rightarrow\rangle}{\sqrt{2}}|\Psi^+_{\mathbf{nm}}\rangle\, .
\end{align}
In each four-state sector the Hamiltonian may be written as
\begin{align}
\label{eq:Hmu}\hat{H}_{\mu,M_x}&=2M_x\Omega\mathbb{I}+\left(\begin{array}{cccc} U_{\mu,M_x}&0&J&J\\ 0&U_{\mu,M_x}&-J&-J\\ J&-J&0&0\\ J&-J&0&0\end{array}\right)\, ,
\end{align}
where $\mu=t,s$, $U_{t,M_x}=V_{M_x}$, $U_{s,0}=U_0$, $\mathbb{I}$ is the identity operator, and we have set $\Omega_{\mathbf{n}}=\Omega_{\mathbf{m}}=\Omega$ and $J_{\mathbf{n}}=J_{\mathbf{m}}=J$ for simplicity.  The $s,0$ singlet sector is rigorously decoupled from the others, but there is mixing between the triplet sectors proportional to the spin model parameters $\chi$ and $C$.  These couplings are neglected due to the large single-particle energy difference $\sim \Omega$ between coupled manifolds with the assumed separation of energy scales $\Omega\gg J,V,U$.

The Hamiltonians Eq.~\eqref{eq:Hmu} may be readily diagonalized, leading to the eigenvalues $2M_x\Omega,2M_x\Omega+U,2M_x\Omega+\frac{U}{2}\pm \sqrt{4J^2+\frac{U^2}{4}}$.  At high temperatures compared to the tunneling, and for the adiabatic preparation procedure described in the main text, the observed dynamics will be an equal weight superposition of the dynamics from the  initial states $|2^{++},0\rangle$, $|0,2^{--}\rangle$, $|\rightarrow_{\mathbf{n}},\leftarrow_{\mathbf{m}}\rangle$, and $|\rightarrow_{\mathbf{m}},\leftarrow_{\mathbf{n}}\rangle$.  The dynamics of the number of excitations following adiabatic conversion back to the ``bare" $g/e$ basis for these states are
\begin{align}
|2^{++},0\rangle,|0,2^{--}\rangle\to n_e\left(t\right)=&1-\cos\frac{V_1 t}{2}\cos\left(t \sqrt{4J^2+\frac{V_1^2}{4}}\right)-\frac{V_1\sin\frac{V_1 t}{2}\sin\left(t\sqrt{4J^2+V_1^2/4}\right)}{\sqrt{16J^2+V_1^2}}\, ,\\
|\rightarrow_{\mathbf{n}},\leftarrow_{\mathbf{m}}\rangle,|\rightarrow_{\mathbf{m}},\leftarrow_{\mathbf{n}}\rangle\to n_e\left(t\right)=&1-\frac{1}{2}\left(\cos\frac{V_{0} t}{2}\cos\left(t \sqrt{4J^2+\frac{V_{0}^2}{4}}\right)+\cos\frac{U_{0} t}{2}\cos\left(t \sqrt{4J^2+\frac{U_{0}^2}{4}}\right)\right)\\
\nonumber &-\frac{1}{2}\left(\frac{V_{0}\sin\frac{V_{0} t}{2}\sin\left(t\sqrt{4J^2+V_{0}^2/4}\right)}{\sqrt{16J^2+V_{0}^2}}+\frac{U_{0}\sin\frac{U_{0} t}{2}\sin\left(t\sqrt{4J^2+U_{0}^2/4}\right)}{\sqrt{16J^2+U_{0}^2}}\right)\, .
\end{align}
For small interactions compared to tunneling, we can expand the result to lowest order in interactions, and find the thermally averaged dynamics
\begin{align}
n_e\left(t\right)&=2\sin^2\left(J t\right)-\frac{\left(U_0^2+V_0^2+2V_1^2\right)t\left(\sin\left(2 Jt\right)-2J t\cos\left(2 J t\right)\right)}{64 J}\, .
\end{align}

In addition to the above two-particle dynamics, there will be contributions from experimental realizations in which the resonant double well has only a single particle.  Here, the dynamics is $n_e\left(t\right)=\sin^2\left(J t\right)$.  Writing $p_1$ as the probability of a single particle in a given double well and $p_2$ as the probability of having two particles, the small-interaction dynamics averaged over all realizations is hence
\begin{align}
n_e\left(t\right)&=\left(p_1+2p_2\right)\sin^2\left(J t\right)-p_2\frac{\left(U_0^2+V_0^2+2V_1^2\right)t\left(\sin\left(2 Jt\right)-2J t\cos\left(2 J t\right)\right)}{64 J}\, .
\end{align}
Noting that $p_1+2p_2=N$ is the average number, we can subtract the measurements of $n_e\left(t\right)/N$ for two different densities with single- and double-occupancy probabilities $(p_1,p_2)$ and $(p_1',p_2')$ and numbers $N$ and $N'$, respectively, to find
\begin{align}
\Delta [n_e\left(t\right)/N]&=-\left(\frac{p_2}{N}-\frac{p_2'}{N'}\right)\frac{\left(U_0^2+V_0^2+2V_1^2\right)t\left(\sin\left(2 Jt\right)-2J t\cos\left(2 J t\right)\right)}{64 J}\, .
\end{align}
In this way we separate the single-particle contrast decay due to a thermal spread in $J_{\mathbf{n}}$ from the contrast decay due to interactions.

\end{document}